\documentclass[11pt,a4paper]{article}
\usepackage[utf8]{inputenc}
\usepackage[T1]{fontenc}
\usepackage{mathptmx} 
\usepackage[pdftex]{graphicx} 
\usepackage{geometry}
\usepackage[english]{babel} 
\usepackage{calc} 
\usepackage{enumitem} 
\usepackage[]{biblatex}
\usepackage{amsmath}
\usepackage{amsfonts}
\usepackage{amssymb}
\usepackage{xspace}
\usepackage{lmodern}
\usepackage{algorithm}
\usepackage{algpseudocode}
\usepackage{rotating}
\usepackage{tikz}
\usepackage{placeins}
\usetikzlibrary{decorations}
\usepackage{csquotes}
\usepackage{caption}
\usepackage{subcaption}

\usepackage{graphicx}
\usepackage{subcaption}
\usepackage{subcaption}

\usepackage{tabularx}
\usepackage{caption}

\usepackage[pdftex,linkcolor=black,pdfborder={0 0 0}]{hyperref}
\usepackage{hyperref}
\usepackage{cleveref}
\usepackage{caption}
\setlist[enumerate]{itemsep=0mm}
\usepackage{orcidlink}
\newcommand{\Ex}{\mathbb{E}}

\newcommand{\Dpel}{$\Delta P\&L\,\,$}

\DeclareMathOperator*{\argmax}{arg\,max}
\DeclareMathOperator*{\argmin}{arg\,min}

\usepackage{academicons}
\usepackage{xcolor}
\def\orcid#1{\kern .08em\href{https://orcid.org/#1}{\includegraphics[keepaspectratio,width=0.7em]{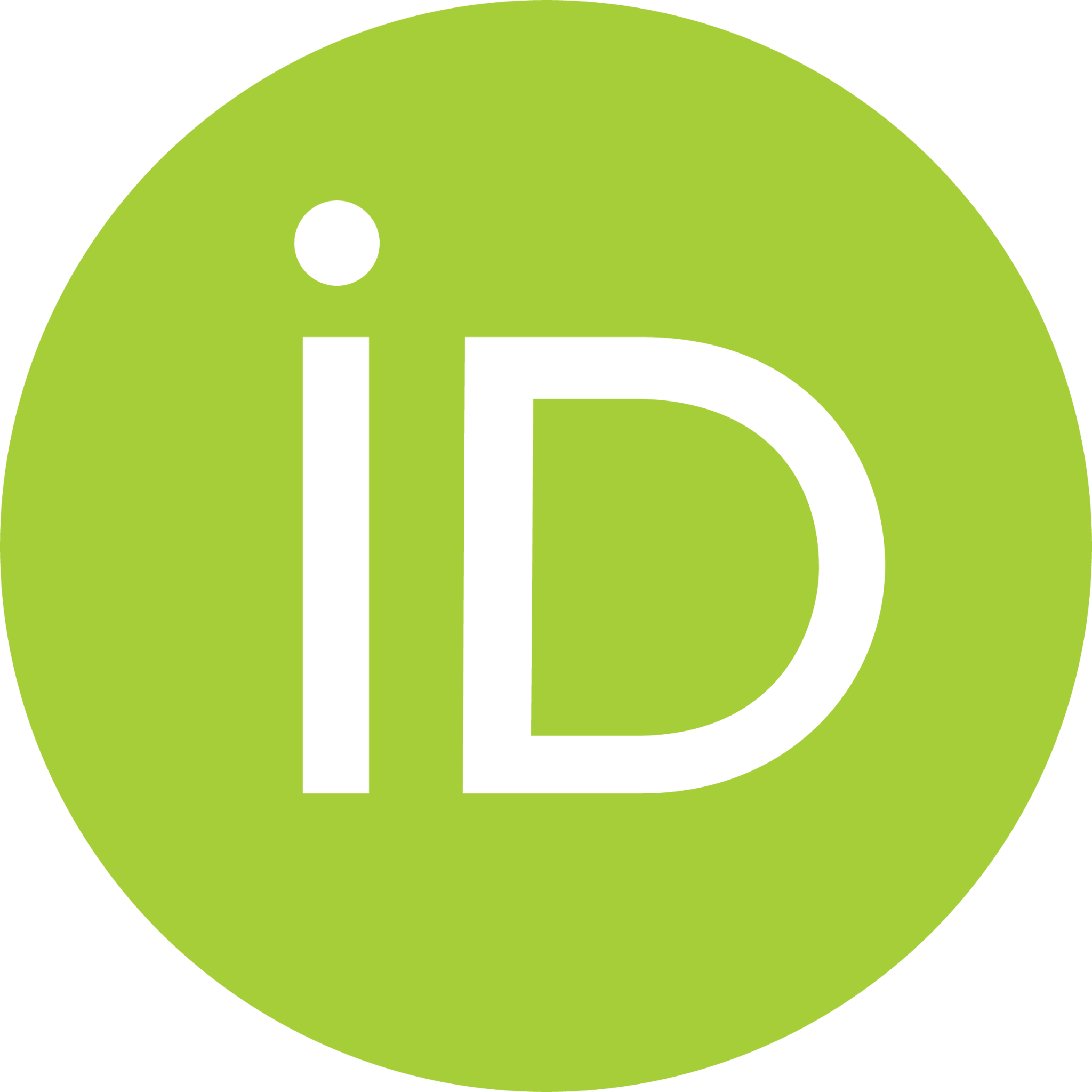}}}
\hypersetup{
  pdfsubject = {OptimalExecutionDDQN},
  pdftitle = {DDQNOptimalExecution},
  pdfauthor = {Andrea Macrì}
}

\begin{document}

\title{Reinforcement Learning for Optimal Execution \\ when Liquidity is Time-Varying }
\author{Andrea Macrì$^1$\footnote{andrea.macri@sns.it} \orcid{0000-0001-6526-5884} 
and Fabrizio Lillo$^{1,2}$\footnote{fabrizio.lillo@sns.it}\\
\small{$^1$ Scuola Normale Superiore, Pisa, Italy}\\
\small{$^2$ Dipartimento di Matematica University of Bologna, Bologna, Italy}}

\date{\today}
\maketitle
\begin{abstract}
 Optimal execution is an important problem faced by any trader. Most solutions are based on the assumption of constant market impact, while liquidity is known to be dynamic. Moreover, models with time-varying liquidity typically assume that it is observable, despite the fact that, in reality, it is latent and hard to measure in real time. In this paper we show that the use of Double Deep Q-learning, a form of Reinforcement Learning based on neural networks, is able to learn optimal trading policies when liquidity is time-varying. Specifically, we consider an Almgren-Chriss framework with temporary and permanent impact parameters following several deterministic and stochastic dynamics. Using extensive numerical experiments, we show that the trained algorithm learns the optimal policy when the analytical solution is available, and overcomes benchmarks and approximated solutions when the solution is not available.
\end{abstract}

\section{Introduction}
\label{intro}
The problem of optimally executing financial trades has been extensively studied for at least two decades. Building on the seminal papers of Bertsimas and Lo (1998) \cite{bertsimas1998optimal} and Almgren and Chriss (2000) \cite{almgren}, the concept of dividing large sell (buy) orders into smaller trades, considering risks and rewards, has been widely applied and extended through mathematical optimisation. Starting with the seminal, yet simplistic, Almgren-Chriss setting, which assumes an Arithmetic Brownian motion for mid-price movements with linear permanent and temporary price impacts, the literature has progressed to more complex models, notable examples are \cite{gueant2015general,gueant2012optimal,cartea2016incorporating,casgrain2019trading,cartea2015algorithmic,bouchaud2009markets,bouchaud2003fluctuations,gatheral2012transient,obizhaeva2013optimal}. 
Since these approaches make use of specific market impact models, which might or might not accurately describe real data, the non-parametric approach based on  Reinforcement Learning (RL) has increasingly been used to solve optimal liquidation problems. As described by \cite{sutton1998reinforcement}, RL models an agent that sequentially learns optimal policies in a generalised environment, optimising a reward function with respect to future discounted cumulative rewards based on actions taken at each time step. Q-learning is one of the most widely used RL algorithms. 
Q-learning in tabular form, was firstly applied by \cite{nevmyvaka2006reinforcement} in the context of optimal execution. In their setting a tabular Q-agent has to decide how many stocks to sell within a certain time window using limit orders. The amount and the execution price at which these stocks are sold depend on the time left for execution, on the inventory left to be executed, and on market prices. Ref.
\cite{hendricks2014reinforcement} adopts a hybrid approach, both analytical and RL-based, where, starting from the optimal execution à la Almgren-Chriss, a stylised agent, still modelled with a classic tabular Q-Learning algorithm, is able to modify its trade execution schedule to minimise a reward function that corresponds to the Implementation Shortfall. 
\par More recently, due to the recent advancements in machine learning methods, the lookup table has been replaced by deep neural nets, thus Deep Q-Learning (DQL) has been widely applied to various financial problems, including market making \cite{abernethy2013adaptive,kumar2020deep}, portfolio optimisation \cite{yu2019model}, trading \cite{kearns2013machine,wei2019model,briola2021deep,jaisson2022deep} and optimal trade execution \cite{ning2021double,daberius2019deep,karpe2020multi,jeong2019improving,schnaubelt2022deep}.\footnote{For a more comprehensive overview on the state of the art on RL methods in finance please refer to \cite{sun2023reinforcement} and \cite{hambly2021recent}}
For what concerns optimal execution problems, a major recent contribution is the paper by Ning et al. (2021) \cite{ning2021double}, where a Double Deep Q-Network (DDQN) algorithm for optimal execution is introduced, overcoming the bias in Q-value estimation found in traditional DQL. This technique, incorporating two neural networks (Q-network and target Q-network), improves policy estimation accuracy. Ref. \cite{ning2021double} considers a Almgren-Chriss framework without permanent impact and with a constant (across time and stocks) temporary impact. 
When applied to real data, they found that DDQN outperforms the Time-Weighted Average Price (TWAP) strategy in seven out of nine stocks considered.
Other notable contributions include \cite{jeong2019improving}, where DQL devises a trading strategy tested on various stock indexes, and \cite{daberius2019deep}, comparing DDQN and Proximal Policy Optimisation (PPO) algorithms on artificial data. PPO and DDQN outperform TWAP when it is not optimal and perform similarly to a slice-and-dice strategy when it is.
\par
\smallskip
 Our paper considers the application of RL for optimal execution to a more realistic environment. In particular, we consider a setting where liquidity, as described by the temporary and permanent impact coefficient, are not constant but time varying. It is well known that liquidity is a dynamic and latent variable. Some papers (for example \cite{barger2019optimal} and \cite{fouque2022optimal}) have considered optimal execution with time varying liquidity, 
 but the underlying assumption is that the model's parameters are known, while their real time estimation is typically very complicated and noisy as shown in for instance in \cite{campigli2022measuring} and \cite{mertens2022liquidity}. 
 Thus estimating the impact parameters and simultaneously adjusting the trading strategy is very challenging and here, in a simulated environment, we test if RL can be successfully used. To this end, we consider an Almgren-Chriss market where the permanent and temporary coefficients are linear but time-varying. We consider different dynamics, from very simple and deterministic ones to the case when they follow a stochastic process. We find that when a closed form solution is known, the RL algorithm finds very similar strategies with similar costs. More interestingly, when the solution is not known or is known only in some regime parameters of low variability, the RL algorithm outperforms the benchmarks.
In summary, the aim is to train a `model robust' agent capable of adapting its trading strategy based on the current liquidity profiles in the market. This entails leveraging minimal information, represented as features to the neural networks, to agnostically select the correct model and selling schedule to follow.\par
\smallskip
The paper is organised as follows: Section 2 introduces baseline models and the DDQL algorithm, Section 3 discusses the main results, and finally Section 4 provides conclusions and outlines further research directions.
\section{Methods}
\label{sec:finMod}
We will consider the problem of a trader who wants to unwind a portfolio of $q_0$ stocks within a time window $[0,T]$.
The goal is to unwind the portfolio \textit{optimally} using arrival price as benchmark, i.e. minimising the difference between the initial and final portfolio values. We assume that the trading influences the stock price via permanent and temporary impacts. The magnitude and dynamics of these impacts will change over the trading experiments that will be presented.
Initially, we assume price dynamics to follow the classic Almgren-Chriss \cite{almgren} model, and then we test these results against those obtained by an agent modelled with a DDQL algorithm. We will further modify and complicate market dynamics to test RL's capabilities in more complex environments. This section introduces the main tools used in the experiments, namely baseline financial models and the DDQL algorithm used to model our agent.
\subsection{Market impact models}
\label{financial models}
\paragraph{Baseline Model}
The Almgren and Chriss \cite{almgren} (A\&C) model is used as the baseline model. We assume a trader with an initial inventory \(q_0\) stocks, the trader needs to unwind this portfolio within a time window \([0,T]\), divided into \(N\) sub-intervals of length \(\tau = \frac{T}{N}\). Setting $t=1,2,...,N$ the model reads:
\begin{align}
\begin{split}
    \label{perm}
    {S}_{t} &= S_{t-1} - g\left( \frac{v_t}{\tau}\right)\tau+\sigma\tau^{\frac{1}{2}}\xi\\
    \Tilde{S_t} &=  {S}_{t-1} - h\left( \frac{v_t}{\tau}\right)
\end{split}
\end{align}
where \(S_t\) is the mid-price for the stock at time $t$. $S_t$ evolves  because of a diffusion part \(\xi\) (a draw from a standard normal random variable) multiplied by the price volatility \(\sigma\). The number of shares sold during the interval $[t-1,t]$ of length \(\tau\) is \(v_t\). The mid-price \(S_t\) is impacted by the \emph{permanent impact} term \(g_t\left( \frac{v_t}{\tau}\right)\), assumed to be linear and constant: \(g\left( \frac{v_t}{\tau}\right) = \kappa \frac{v_t}{\tau}\). The price received by the trader, \(\Tilde{S_t}\), is equal to the mid-price impacted by a \emph{temporary impact} term also assumed to be linear and constant in time: \(h(\frac{v_t}{\tau}) = \Tilde{\alpha}\frac{v_t}{\tau}\), in what follows we use and model directly the quantity \(\alpha = \frac{\Tilde{\alpha}}{\tau}\). 
The aim is to find an optimal quantity to sell \(v_t\) for each sub-interval \(t\), minimising the \emph{Implementation Shortfall} (IS) cost functional given by:
\begin{equation}
\label{is}
    IS = S_0q_0 - \sum^N_{t=1} \Tilde{S_t}v_t
\end{equation}
Since $IS$ is stochastic, A\&C assumes a mean variance optimisation (or equivalently a CARA utility maximiser) characterised by a risk aversion parameter $\lambda$. A\&C shows that the optimal inventory holding $q_t$ for each time \(t=1,\dots N\) is:
\begin{equation*}
    q^*_t = q_0\frac{\sinh({\omega(T-t)})}{\sinh{(\omega T)}}
\end{equation*}
where \(\omega\) solves \(2(\cosh{(\omega\tau}) - 1) = \frac{\lambda\sigma^2}{2\Tilde{\alpha}}\tau^2\).

In this paper the trader (or agent) is assumed to be risk neutral with \(\lambda=0\), thus a \emph{risk-neutral agent} in discrete time adopts the TWAP strategy, selling their inventory at a constant rate \(v^* = \left( \frac{q_0}{N}, \dots , \frac{q_0}{N}  \right)\) or equivalently holding \(q^*_t = (N-t)\frac{q_0}{N}\) stocks for each time \(t=1,\dots N\).

\paragraph{Time-dependent impacts baseline model}
We then consider a different baseline model by assuming that the permanent and the temporary impacts are \emph{time-varying}. This is a more realistic baseline model assumption, since in real-market situations liquidity is variable and fluctuates. In fact, as discussed in \cite{campigli2022measuring}, price impacts have strong intraday time-dependency since they react quite quickly to changing market conditions. 
Impact dynamics usually have a deterministic and a stochastic component. In this baseline we consider deterministic dynamics, while in the next one we will consider a stochastic dynamics for the impacts.
\par
In general, if both impacts are time-varying, the functional to be minimised is similar to the one on the right-hand side of \eqref{is} but with time-varying impacts, i.e.:
\begin{equation}
\label{tv}
        \sum^N_{t=1} \Tilde{S_t}v_t = \sum^N_{t=1} (S_{t-2} - \kappa_{t-1}v_{t-1})v_t - \sum^N_{t=1}\alpha_t(v_t)^2
\end{equation}
In our context, the time varying permanent impact is \(g(\frac{v_t}{\tau}) = \kappa_t\frac{v_t}{\tau} \). We consider here the simplest deterministic dynamics for $\kappa_t$, namely we consider a linear temporal dependence, \(\kappa_t = \kappa_0 \pm \beta_{\kappa}\times t \). The time varying temporary impact is assumed to be similarly specified, i.e., \(h(v_t) = \alpha_t v_t\),  with dynamics \(\alpha_t = \alpha_0 \pm \beta_{\alpha} \times t\). 
Since impacts must be positive, we choose the parameters in such a way that both $\kappa_t>0$ and $\alpha_t>0$, $\forall t$. Fig. \ref{fig:imp_tv} below shows the impact parameter dynamics used in our experiments.
\par
The optimal schedule in this scenario might be found via standard quadratic optimisation techniques. However, when the dynamics of the impacts are not known, blindly applying optimisation methods may lead to inefficient trading schedules. We test the DDQL results against those obtained via canonical methods as if the dynamics were known from the beginning. The aim is to test whether the algorithm has learnt the correct parameter specification for both the impacts without prior knowledge of their deterministic dynamics, in a quest to obtain a `robust' optimal execution schedule in an agnostic way.
\paragraph{Stochastic impacts baseline model}
Lastly, we consider an environment with stochastic impacts. The use of both permanent and temporary stochastic impacts allows for more freedom when it comes to model liquidity. Thus, both the temporary and the permanent impacts are assumed to follow a square-root mean reverting process.
We will choose the parameters in such a way that the impacts are almost surely positive, we will further assume positive correlation between the permanent and the temporary impacts. 
The stochastic impacts model was proposed Barger and Lorig (2019) \cite{barger2019optimal}, where an \emph{approximate} optimal execution solution is found via a Taylor approximation around the long run mean values for the processes considered.
According to \cite{barger2019optimal}, the model is:
\begin{align}
\begin{split}
    \label{stochImp}
    dS_t &= -\kappa_t\nu_tdt+\sigma dW_t\\
    d\kappa_t &= \lambda_\kappa(\theta_\kappa - \kappa_t)dt + \sigma_\kappa\sqrt{\kappa_t}dB^{(1)}_t\\
    \Tilde{S_t} &= S_{t-1} - \alpha_t\nu_t\\
    d\alpha_t &= \lambda_\alpha(\theta_\alpha  - \alpha_t)dt + \sigma_\alpha\sqrt{\alpha_t}dB^{(2)}_t\\
    <&B^{(1)}_t, B^{(2)}_t> = \omega
\end{split}
\end{align}
where $S_t$ is the stock price, $\Tilde{S_t}$ is the execution price, $\lambda_\kappa$ and $\theta_\kappa$ are, respectively, the rate of mean reversion and the long run mean at which the permanent impact reverts to, $\lambda_\alpha$ and $\theta_\alpha$ have same meaning but for the temporary impact dynamics. \par
At time $t$, the trading velocity $\nu_t$, that is the instantaneous trading volume for a sell program is found, assuming that $\alpha_0 = \theta_\alpha,\,\, \kappa_0 =\theta_\kappa$, as:
\begin{equation}
\label{actStoch}
    \nu_t = \left( \frac{1}{N-t} + \frac{\lambda_\alpha(\theta_\alpha  - \alpha_t)}{2\alpha_t} + (N-t) \frac{\lambda_\kappa(\theta_\kappa  - \kappa_t)}{6\kappa_t}\right) \times q_{t}
\end{equation}
Eq.~\eqref{actStoch} can be seen as a perturbation of the TWAP solution, which is optimal in the constant impact case. 
In what follows we will consider a discrete-time version of the model. We will assume that the trader trades within each of the $N$ time-steps $t$, a quantity $v_t = \nu_t\tau$ of stocks.
The price and the impacts' paths over the $N$ discrete time-steps are simulated using the model in Eq.~\eqref{stochImp}.

\subsection{DDQL algorithm}
Deep Reinforcement Learning is a machine learning paradigm where an agent, modelled with neural networks, learns how to interact with an environment. This is achieved by taking actions and receiving feedback in the form of rewards or penalties. The primary objective of the agent is to maximise the cumulative reward over time by discovering an effective policy or strategy.
\par
We denote the sets of all possible states, actions, rewards, and policies as $\mathcal{S}$, $\mathcal{A}$, $\mathcal{R}$ and $\Pi$ respectively. In this context, the algorithm, given the environment's state $s_t \in \mathcal{S}$, must decide which action $a_t \in \mathcal{A}$ to perform. This action modifies the environment to a subsequent state $s_{t+1} \in \mathcal{S}$ and results in receiving a reward $r_{t} \in \mathcal{R}$. The ultimate goal is to learn an optimal policy $\pi^* \in \Pi$ that maximises the sum of discounted future rewards, denoted as $\sum^{\infty}_{t=0} \gamma^k r_{t}$, where $\gamma \in (0,1]$ is the discount factor.
\par
\par
In this paper, we employ Deep Reinforcement Learning to find the optimal execution schedule in various time-varying impact settings. Specifically, we adopt the \emph{Double Deep Q-Learning} (DDQL) algorithm, providing a model-agnostic solution to the optimal execution problem. Q-values, representing the quality of taking actions in each state of the environment, are obtained through neural networks known as Q-nets. We use Double Deep Q-learning, thus employing two neural networks, namely the main Q-net ($Q_{\text{main}}$) for action selection and the target Q-net ($Q_{\text{tgt}}$) for state evaluation.
Overall, the algorithm undergoes an exploration phase, where it explores possible rewards by taking random actions in evolving states. Subsequently, it gradually transitions to an exploitation phase, leveraging learned information on rewards using $Q_{\text{main}}$ to select actions. 
To maintain stability, the weights for $Q_{\text{tgt}}$ are periodically synchronised with those of $Q_{\text{main}}$. 

This is achieved through a training routine that updates the main net using past states, actions, and rewards randomly sampled from a memory.
The training occurs over $M$ episodes, followed by a test phase on an additional $B$ episodes to evaluate the DDQL agent's performance post-exploration.

\subsubsection{Algorithm setup for numerical experiments}

In order to tackle the problem of optimal execution with DDQL, we consider an agent who wants to sell a portfolio of $q_0$ stocks within a time-window $[0,T]$. During this time window, the agent totally sells its inventory and this is called an \emph{episode}. The episodes are assumed to be divided into $N$ time-steps and the time increment for the simulations is $\tau = \frac{T}{N}$. In this context, the agent decides the quantity of stocks $v_t$ to sell in each of the $t$ time-steps in which the episode is divided. The mid-price at which the agent trades is modeled according to Eq.~\eqref{perm}, but depending on the frameworks considered, the dynamics for the impact parameters change.\par 
Over the $M$ train episodes the algorithm learns, via exploration-exploitation,  the best trading strategy and changes the weights of the Q-nets accordingly. The architecture of the algorithm in the training phase is thus divided into two phases: an \emph{exploration} and an \emph{exploitation} phase, managed by the parameter $\epsilon \in (0,1]$ that will decrease during training and initially set as $\epsilon=1$. Depending on the phase, the way actions $v_t$ are chosen changes.  When the agent is exploring, it will randomly select sell actions $v_t$ in order to explore different states and rewards in the environment, alternatively, when the agent is exploiting, it will use $Q_{\text{main}}$ to select $v_t$.

\paragraph{Action selection and reward function.}

At the beginning of each time interval and for each episode, the agent has limited knowledge of the environment where it trades. This means that it has access to states $s_t$ made either by tuples of ($q_t,t$) or ($q_t,t, S_{t-1}$) (i.e. quantity, time and possibly mid-price), depending on the considered feature setup.
\par
Given the current value of $\epsilon$, a draw $\zeta$ from a uniform distribution determines whether to perform exploration or exploitation. Specifically, with probability $\epsilon$
the agent chooses to explore, and the sell action $v_t$ is sampled from a binomial distribution with number of trials equal to $q_t$ (i.e. the inventory left at the beginning of a sub-interval) and probability of success is $\frac{1}{N - t}$. In this way, in the exploration phase the TWAP is selected on average.
Alternatively (with probability $1-\epsilon$), the agent chooses the Q-optimal action, i.e. the action that maximises the Q-value from $Q_{\text{main}}$, doing exploitation of what learnt in the exploration phase. Notice that the agent cannot sell more than the remaining inventory and, in addition, no buy actions can be performed in the selling program.\par
In this way, the agent randomly explores a large number of states and possible actions to perform.
Once every $m$ actions taken during training episodes, $\epsilon$ is multiplied by a constant $c<1$ such that $\epsilon \gets \epsilon \times c$, in this way for large number of episodes $M$, $\epsilon$ converges to zero and the algorithm gradually stops exploring and starts to greedily exploit what it has learned in terms of weights $Q_{\text{main}}$.
In formulae, the action decision rule for each time step $t$  in the training phase is:
\begin{align}
    \begin{split}
        \label{eq:action}
        &\epsilon\in(0,1)\,\,,\,\, \zeta\sim\mathcal{U}(0,1)\\
        &v_t = 
        \begin{cases}
            \sim \text{Bin}(q_{t},\frac{1}{N-t})\,\,\,\,\,\,\,\,\,\,\,\,\,\,\,\,, \text{if}\,\,\zeta \leq \epsilon\\
            \argmax_{v'\in[0,q_{t}]}Q_{M}(s_{t},v'|\theta_{\text{main}}) \,\,\,\,\,\,\,\,, \text{else}
        \end{cases}
    \end{split}
\end{align}
Subsequently, at each $t$, once that the action is chosen according to Eq.~\eqref{eq:action}, the reward is calculated as follows:
\begin{equation}
\label{eq:reward}
    r_t = S_{t-1} v_t - \alpha v_t^2
\end{equation} 
Overall, the rewards for every $t\in [1,N]$ are:
\begin{align}
\begin{split}
    \label{eq:rew_expanded}
    r_1 &= S_{0} v_1 - \alpha v_1^2\\
    r_2 &= (S_1 + \kappa  v_1 + \mathcal{N}(0,\sigma))v_2 - \alpha  v_2^2\\
    \vdots&\\
    r_{N} &= (S_{N-1} + \kappa  v_{N-1}  + \mathcal{N}(0,\sigma))v_{N} - \alpha  v_{N}^2\\
    \end{split}
\end{align}
Thus, for each time step $t$ the agent sees the reward from selling $v_t$ shares, and stores the state of the environment, $s_t$, where the sell action was chosen along with the reward and the next state $s_{t+1}$ where the environment evolves. As said, the aim of the agent is to cumulatively \emph{minimise} such reward, such that the liquidation value of the inventory happens as close as possible to the initial value of the portfolio. At the end of the episode, the reward per episode
 is $-q_0 S_0 + \sum^N_{t=1} S_{t-1}v_t - \alpha v^2_t$.
 \paragraph{Training scheme}
The states $s_t$, actions $v_t$, rewards $r_t$ and subsequent future states $s_{t+1}$ obtained by selling a quantity $v_t$ in state $s_t$, form a \emph{transition} tuple that is stored into a memory of maximum length $L$. As soon as the memory contains at least  $b$ transitions, the algorithm starts to train the Q-nets. In order to train the nets, the algorithm samples random batches of length $b$ from the memory, and for each sampled transition $j$ it calculates:
\begin{align}
\label{y_T}
y^j_{t}(\theta_{\text{tgt}}) = 
                    \begin{cases}
                        r^j_{t} & \text{if } {t}=N;\\
                        r^j_{t} + \gamma Q_{\text{tgt}}(s^j_{t+1},v^*|\theta_{\text{tgt}}) & \text{else}
                    \end{cases}
\end{align}
In Eq.~\eqref{y_T}, $r^j_t$ is the reward for the time step considered in the transition $j$, $s^j_{t+1}$ is the subsequent state reached in $t+1$, known since it is stored in the same transition $j$, while $v^* = \argmax_{v}Q_{\text{main}}(s^j_{t},v|\theta_{\text{main}})$. $\gamma$ is a discount factor that accounts for risk-aversion\footnote{We set a $\gamma=1$ since we model a risk neutral agent}. 
The loss to be minimised is the mean squared error between the target $y(\theta_{\text{tgt}})$ and the values obtained via the Q-main network. Thus, it is: 
$$    \mathcal{L}(\theta_{\text{main}},\theta_{\text{tgt}}) = 
                    \frac{1}{b} \sum^b_{\ell =1}\left( \left[  y^\ell_t(\theta_{\text{tgt}}) - 
                    Q_{\text{main}}(s^\ell_t,v^\ell_{t}|\theta_{\text{main}}) \right]\right) ^2$$
$$\theta^*_{\text{main}} = \argmin_{\theta_{\text{main}} }\mathcal{L}(\theta_{\text{main}},\theta_{\text{tgt}}) $$

The algorithm then uses back propagation and gradient descent to update the weights of the main network. This sub-routine is repeated for each random batch of transitions sampled from the memory. Finally, as stated above, once every $m$ actions $\epsilon$ is decreased by a factor $c<1$ and the algorithm sets $Q_{\text{tgt}}= Q_{\text{main}}$.\par 
After training, other $B<M$ trading episodes are fed to the algorithm in order to exploit what learnt during training, this is the testing phase and the actions are now selected using just $Q_{\text{main}}$. The results shown below are those obtained in the test phase only.\par
The features of the Q-nets are ($q_t,t,v_t$) or ($q_t,t, S_{t-1},v_t$) and are normalised in the domain $[-1,1]$ using the procedure suggested in \cite{ning2021double}, normalised mid-prices $\Bar{S}_t$ are obtained via min-max normalisation.
In our setup we use fully connected feed-forward neural networks with $5$ layers, each with $30$ hidden nodes 
activation functions are leakyReLu, and finally we use the ADAM optimiser for optimisation. The parameters used to calibrate the algorithm are reported in Table~\ref{tab:pmts}, the training algorithm is reported in Algorithm~\ref{alg:cap}.
\FloatBarrier
\begin{table}[ht]
  \centering
  \caption{Fixed parameters used in the DDQL algorithm. The parameters not shown in the table change depending on the experiments and are reported accordingly.}
  \label{tab:pmts}
  \begin{tabularx}{\textwidth}{|X|X|X|X||X|X|}
    \hline
    \multicolumn{4}{|l||}{\textbf{DDQL parameters}} & \multicolumn{2}{l|}{\textbf{Model parameters}} \\
    \hline
    NN layers & $5$               & $M$ train eps.  & $10.000$ & $N$ intervals     & $10$ \\
    Hidden nodes & $30$           & $B$ test eps.   & $5.000$ & $\sigma$-stock    & $0.00001$ \\
    ADAM lr & $0.0001$   & $m$ reset rate & $100$ acts.  & $S_0$-stock       & $10\$$ \\
    Batch size $b$ & $32$         & $c$ decay rate & $0.995$ & $q_0$ inventory   & $20$ \\
    $L$ memory length & $15^.000$ & $\gamma$-discount & $1$  &                   & \\
    \hline
  \end{tabularx}
\end{table}
\FloatBarrier

\FloatBarrier
\begin{algorithm}[ht]
\caption{Training of Double Deep Q-Learning agent for optimal execution}
\begin{algorithmic}
\Require
{\\

Initialise with random weights $Q_{\text{main}}$ and make a copy $Q_{\text{tgt}}$;\\
Initialise the memory with max length $L$. Set $\epsilon = 1$, $b$ batch size, $M$ train episodes;\\
Set market dynamics parameters;\\
}
\For {i in M}
    \State Set $S^i_0 = S_0$;\\
    \For{t in N}
        \State {$s_{t} \gets (q_t,t)$;}  \Comment{or $(q_t, t, S^i_{t-1})$ if mid-price considered}
        \State {$v_t \gets  \begin{cases}
            \text{sample\,\,}\text{Bin}(q_{t},\frac{1}{N-t})\,\, \text{with probability}\,\,\epsilon\\
            \argmax_{v'\in[0,q_t]}Q_{\text{main}}(s_{t},v'|\theta_{\text{main}}) \,\, \text{with probability}\,\,(1 -\epsilon)
            \\
        \end{cases};$}\\
        \State{$r_t\gets S^i_{t-1} v_t - \alpha v^2_t$;}\\
        \State{$S^i_{t-1}\to S^i_t$;}\Comment{Generate $S^i_t$ from $S^i_{t-1}$}\\
        \State{$s_{t+1} \gets (q_{t+1}, t+1)$ ;}\Comment{or $(q_{t+1}, t+1, S^i_{t})$ }\\
        \State {Memory $\gets(s_{t},r_{t},v_t, s_{t+1})$;} \Comment{Memory storing}
        \\
        \If{Length of memory $\ge$ $b$}
            \For{j in $b$}
                \State{Sample a batch of $(s^j_{t},r^j_{t},v^j_t,s^j_{t+1})$ from memory;}
                \State{$v^* = \argmax_{v}Q_{\text{main}}(s^j_t,v|\theta_{\text{main}})$;}
                \State{$y^j_{t}(\theta_{\text{tgt}}) = 
                    \begin{cases}
                        r^j_{t} & \text{if } {t}=N;\\
                        r^j_{t} + \gamma Q_{\text{tgt}}(s^j_{t+1},v^*|\theta_{\text{tgt}}) & \text{else}
                    \end{cases}$} 
            \EndFor\\
            \State{$\theta^*_{\text{main}} = \argmin_{\theta_{\text{main}}}\mathcal{L}(\theta_{\text{main}},\theta_{\text{tgt}})$ via gradient descent with loss to minimise:
                    $$\mathcal{L}(\theta_{\text{main}},\theta_{\text{tgt}}) = 
                    \frac{1}{b} \sum^b_{\ell =1}\left( \left[  y^\ell_t(\theta_{\text{tgt}}) - 
                    Q_{\text{main}}(s^\ell_t,v^\ell_{t}|\theta_{\text{main}}) \right]\right) ^2$$}
            \If{Length of memory = $L$} halve the length of memory\EndIf
        \EndIf
        \State{After $m$ iterations decay $\epsilon=\epsilon\times c$;}
        \State{After $m$ iterations $\theta_{\text{tgt}}\gets\theta_{\text{main}}$;}
    \EndFor
\EndFor
\end{algorithmic}
\label{alg:cap}
\end{algorithm}
\FloatBarrier
\section{Experiments and results}

In this section we present the results of our numerical investigations. The aim of the experiments is to study whether the DDQL agent is able to find model-robust optimal execution strategies without any form of knowledge of the underlying impact model, but just based on simple sets of information. Since the algorithm does not know neither the parameters nor the model used for the dynamics of the impacts we adopt a model agnostic approach to find optimal execution strategies.
In this context, we study whether the DDQL agent is able to learn the optimal strategy (i) when the optimal solution is known but the environment is not known to the agent (the impact can be equally increasing or decreasing) and (ii) when the optimal solution is not known in closed form but in an approximate form and the environment is still not known to the agent. In this sense, we investigate the use DDQL as a tool to develop `robust' strategies. In a way, the goal is to use DDQL as a robust non-parametric technique, able to agnostically detect price impacts' magnitude and dynamics in the market by using as few information as possible.\par

Notice that the number of trading steps per episode is set to $N=10$ and the agent has to to sell $q_0=20$ shares.
After the test phase is complete, the average IS is calculated for each test episode. Thus, in the test phase we measure for each episode the IS for both the strategies: the DDQN suggested strategy and the optimal one if impacts dynamics and parameters were known from the beginning.
The performance difference between the two approaches, for each test episode, is quantified as:
$$ \Delta P\&L = \frac{\mathcal{C}_{\text{agent}} - \mathcal{C}_{\text{b}}}{\mathcal{C}_{\text{b}}}$$
where $\mathcal{C} = \sum^N_{t=1} v_t\Tilde{S}_t$, the cash generated by the trading. Notice that $\mathcal{C}_{\text{b}}$ is the cash calculated for the baseline model, while $\mathcal{C}_{\text{agent}}$ is the cash obtained using DDQL strategies. In the tables below we report the average values and the standard deviation over the whole test set.
\subsection{Constant impact}
We first consider mid-price dynamics generated according to the A\&C model of Eq. \eqref{perm}. The impacts are constant and the coefficients are set equal to $\kappa=0.001$ and $\alpha=0.002$. 
When the market behaves as in Eq.~\eqref{perm}, the TWAP is the optimal solution,  i.e. the strategy with minimal costs is the one where $q_0$ is sold equally over the time steps. We want to check that the DDQL is able to recover a similar strategy also by comparing the expected cost with the one of a TWAP. This would imply that DDQL understands that the impact parameters are fixed and trades accordingly with minimal discrepancy compared to the closed-form solution's costs. We do not expect that including price as feature should improve the DDQL strategy performance.

\begin{figure}[ht]
    \centering
    
    \begin{subfigure}[ht]{0.5\textwidth}
        \centering
        \includegraphics[scale=0.45]{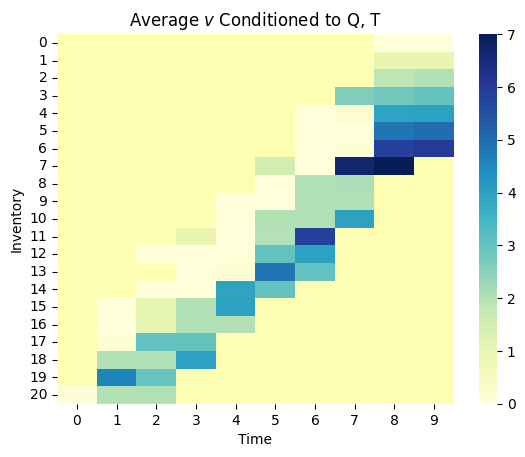}
    \end{subfigure}%
    \begin{subfigure}[ht]{0.5\textwidth}
        \centering
        \includegraphics[scale=0.6]{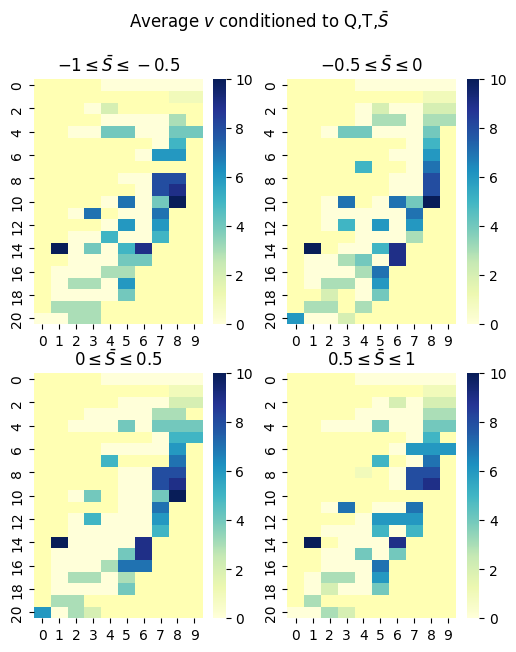}
    \end{subfigure}
    
    \caption{Constant impact. Left panel: Average number of stocks sold per time-step $t$ and inventory level $q_t$. Right panel: Average number of stocks sold per time-step $t$, inventory level $q_t$, and normalised price $\Bar{S}$.}
    \label{fig:AeC_both}
\end{figure}

\paragraph{Features Q,T}
The results with the costs are reported in Table~\ref{tab:AeC}. We notice that, if the agent has knowledge of just $q_t$ and $t$, it retrieves an IS values very similar to the one that would be obtained with the TWAP strategy.
 The \Dpel shows a difference of just $-0.4$ basis points and a standard deviation of only two basis points. The average number of stocks sold per time-step and level of inventory is shown in the left panel of Fig.~\ref{fig:AeC_both}. It can be observed that the agent has learned a strategy consistent with `slice and dice' trading.\par

\paragraph{Features Q,T,S}
If one includes also the price among the features, we observe that the DDQL agent is able to retrieve the same cost. The \Dpel is in the order of $-0.25$ basis points, implying that the optimal cost has been learnt and reproduced by the agent. The average optimal trading per level of inventory $q_t$, time-step $t$, and normalised price level $\Bar{S}$ are reported in the right panel of Fig.~\ref{fig:AeC_both}.\par

The strategy does slightly vary depending on the price level. It might be that during training some states did bear a lower cost. However, the algorithm overall does not perform any better than a noisy TWAP strategy, as expected. Moreover, it appears that allowing for more information in such a simple environment does not significantly add any value. The strategy adjusts depending on the price level considered to be optimal and does not outperform the previous case. This aligns with the theoretical findings in \cite{schied2010optimal}, proving that dynamical strategies (which can depend on price) are also statically optimal.\par
\FloatBarrier
    \begin{table}[htbp]
        \centering
        \begin{tabular}{cccccc}
            \hline
            & \multicolumn{3}{c}{Constant Impact}&  \\
            & \textbf{A\&C} & \textbf{DDQL}  & $\Delta$P\&L &  $\Delta$P\&L \\
            \textbf{features} & $\Ex[IS]$ & $\Ex[IS]$  & \textbf{avg.} &    \textbf{std.dev.}\\
            \hline
            \hline
            & & & & & \\
            Q,T & 0.2607  & 0.2698  &  -0.455 & 2.5\\
            & & & & & \\
            Q,T,S & 0.2607  & 0.2652 &  -0.225 & 1.6\\
            & & & & & \\
            \hline
        \end{tabular}
        \caption{Constant impact. Comparison between the costs of the DDQL agent and of the A\&C model. $\Delta$P\&L values are reported in basis points.}
        \label{tab:AeC}
    \end{table}
\FloatBarrier
\FloatBarrier
\FloatBarrier
\subsection{Deterministically time-varying impact}
Having established that the DDQL agent is able to recover the optimal solution when impacts are constant, we test the algorithm with two different, yet deterministic, impacts patterns. It is widely recognised that liquidity has intraday patterns\footnote{see for example \cite{campigli2022measuring} for a thorough discussion on this topic}
thus a trend in impact parameter is observed in certain parts of the trading day, e.g. in the first part of the trading day or around important market announcements. If one knows in advance the functional form of the dynamics of the impact parameters, the optimal execution is again a minimisation of a cost functional and can be solved via quadratic optimisation (although the well posedness of the problem is far from trivial, see \cite{palmari2024}). However, the DDQL agent has no knowledge either about the existence of a pattern or whether the pattern shows an increasing or decreasing trend and the parameters for the impact have to be implicitly estimated by the agent employing DDQL. In making no assumption on market's dynamics the nature of the problem becomes more involved. \par
As explained in Section~\ref{financial models}, we assume impacts follow a linear deterministic trend.
We consider either an increasing or a decreasing impact, and in order to test the appropriateness of the parameters learnt by the algorithm, we calculate the optimal strategy using quadratic optimisation\footnote{We solved the problem using the \textbf{.optimise()} method in SciPy and applying Quadratic programming to Eq.~\eqref{tv}.} and then we simulate the optimal strategy and we measure its cost. The resulting IS is then compared with the one obtained with the DDQL strategy.  We also use as point of comparison the IS of a TWAP strategy, which is obviously not optimal. This serves as an additional benchmark of the performance of the DDQL agent in addition to the theoretically optimal one.
\FloatBarrier
\begin{figure}[h]
    \centering
    \includegraphics[scale = 0.45]{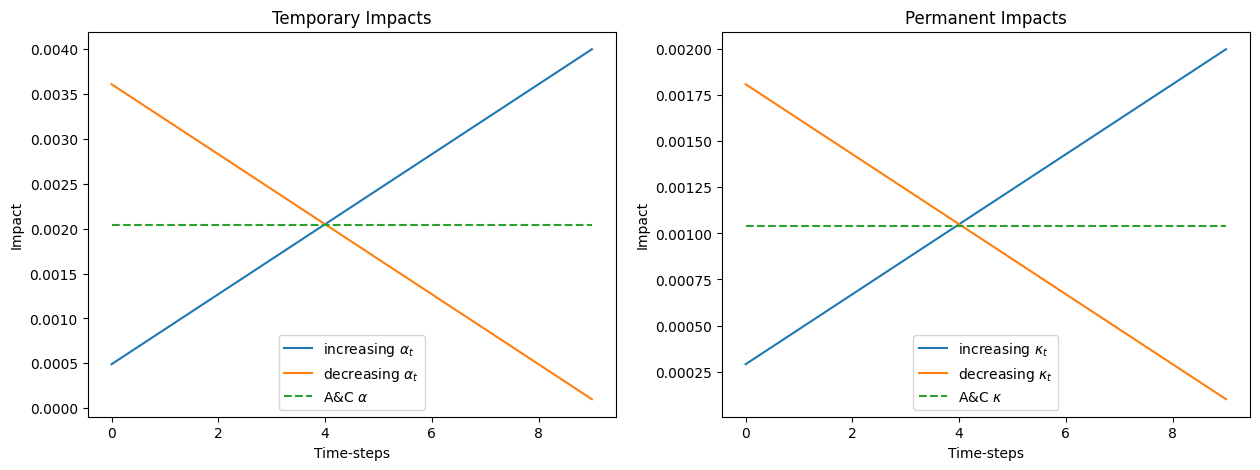}
    \caption{Time-varying dynamics of the temporary and permanent impact}
    \label{fig:imp_tv}
\end{figure}
\FloatBarrier
\subsubsection{Testing and training with increasing impacts}
\begin{figure}[ht]
    \centering
    \includegraphics[scale = 0.45]{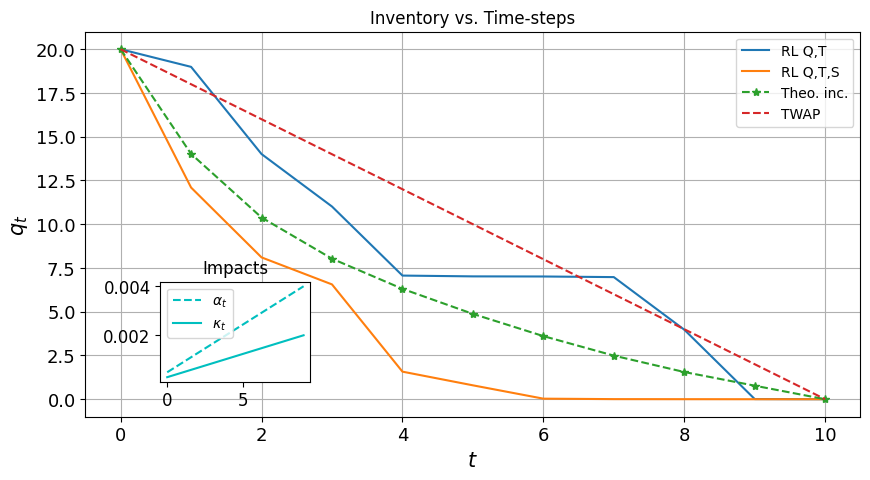}
    \caption{Increasing impact. Average inventory holdings per time step with different features, compared to the optimal inventory holding if the strategy was known from the beginning, and to the inventory holding if the TWAP strategy was used}
    \label{fig:inc_inv}
\end{figure}

The impacts dynamics is $g(\frac{v_t}{\tau}) = \kappa_t \frac{v_t}{\tau} $, where $\kappa_t = \kappa_0 + \beta_{\kappa}\times t $ and $h(v_t) = \alpha_t v_t$,  where $\alpha_t = \alpha_0 + \beta_{\alpha} \times t$ and the parameters for the impacts are reported in Table~\ref{tab:pmts_tv_up}. The parameters are chosen such that the average value of the impact parameter during the trading period are equal to the fixed impacts in the experiments of the previous section. Fig. \ref{fig:inc_inv} shows the optimal solution for these parameters (green line, the red line is the TWAP reference). As expected, given that impacts are increasing, it is better to trade more than in a TWAP at the beginning to exploit the initial lower trading costs.\par
\begin{figure}[ht]
    \centering
    \begin{subfigure}[ht]{0.5\textwidth}
        \centering
        \includegraphics[scale=0.45]{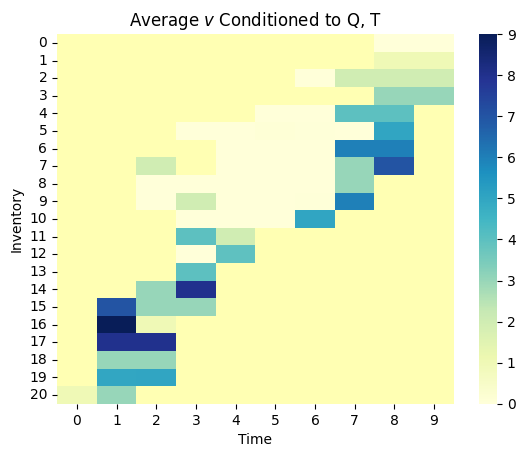}
    \end{subfigure}%
    \begin{subfigure}[ht]{0.5\textwidth}
        \centering
        \includegraphics[scale=0.6]{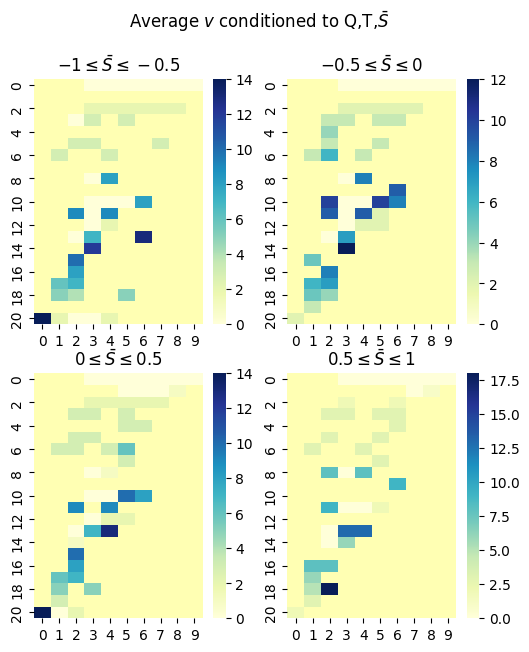}
    \end{subfigure}
    \caption{Increasing impact. Left panel: Average number of shares sold per time-step $t$ and inventory level $q_t$. Right panel: Average number of shares sold per time-step $t$, inventory level $q_t$ and normalised price $\Bar{S}$}
    \label{fig:inc_both}
\end{figure}

\FloatBarrier
\begin{table}[htbp]
  \centering
  \caption{Parameters used in the increasing time varying setting.}
  \label{tab:pmts_tv_up}
    \begin{tabular}{||l|r||}
    \hline
    parameters      & value\\ 
    \hline  
    $\kappa_0$        & 0.0001\\
    $\alpha_0$        & 0.0001\\
    $\beta_{\kappa}$  & 0.0002\\
    $\beta_{\alpha}$  & 0.0004\\
    $\kappa_5=\kappa$ & 0.001\\
    $\alpha_5=\alpha$ & 0.002\\
    \hline
    \end{tabular}
\end{table}
\FloatBarrier
\paragraph{Features Q,T}
As shown in Table~\ref{tab:tv_inc}, when only price and time are used as features,  the DDQL algorithm is not able to really beat the TWAP strategy in terms of costs. In fact, the \Dpel between the DDQL and the TWAP strategy is almost zero, i.e. \Dpel$=-3.20e-5$.
As a consequence, the \Dpel between the DDQL and the theoretical simulated costs is significantly negative (see Table~\ref{tab:tv_inc}). Looking at the blue line in Fig.~\ref{fig:inc_inv} it appears that the DDQL agent with two features learns to trade more at the beginning, but not enough. A similar conclusion is obtained by looking at the full strategy (see left panel of Fig.~\ref{fig:inc_both}).  Looking at the \Dpel value and at the average quantity sold per inventory level and time-step, it can be concluded that market parameters have been noisily estimated leading to a strategy as good as a TWAP strategy.
\paragraph{Features Q,T,S}
Adding the mid-price to the features leads to better results in terms of \Dpel. 
In fact, as reported in Table~\ref{tab:tv_inc} the simulated IS for the DDQL agent performs better than the one obtained with TWAP. 
Comparing our results with the theoretical solution, adding the mid-price as a feature lowers the \Dpel to just $2$ basis points of difference. Although not performing as good as the theoretical solution, adding a feature significantly increases the learning capabilities of the algorithm. This conclusion is supported by the average inventory holding and average trading schedule reported in Fig. \ref{fig:inc_inv} and right panel of Fig.~\ref{fig:inc_both}, respectively. It can be seen that the agent has learnt to trade too aggressively within the first half of the execution window. This is compatible with the increasing nature of the underlying impact parameters. We conclude that the agent has estimated the trend in the impacts and has found a trading strategy accordingly.\par 
\FloatBarrier
\begin{table}[htbp]
    \centering
    \begin{tabular}{ccccc|c}
        \hline
                        & \multicolumn{3}{c}{Increasing Impact}&  \\
                        & \textbf{Theo.} & \textbf{DDQL} & \Dpel vs. Theo.& \Dpel vs. Theo.  & \textbf{TWAP}\\
        \textbf{inputs} & $\Ex[IS]$  & $\Ex[IS]$ & \textbf{avg.}  & \textbf{std.dev.}& $\Ex[IS]$\\
        \hline
        \hline
        & & & & & \\
        Q,T & 0.1449  & 0.2401   & $  $-4.76 & $  $ 1.1 & 0.2326 \\
        & & & & & \\
        Q,T,S & 0.1449  & 0.1944  & $  $-2.42 & $  $ 1.1 & 0.2326\\
        & & & & & \\
        \hline
    \end{tabular}
    \caption{Increasing impact. Comparison between the costs of the DDQL agent, of the theoretically optimal solution, and of a TWAP strategy. $\Delta$P\&L values are reported in basis points.}
    \label{tab:tv_inc}
\end{table}

\FloatBarrier
\subsubsection{Testing and training with decreasing impacts}
\begin{figure}[ht]
    \centering
    \includegraphics[scale = 0.45]{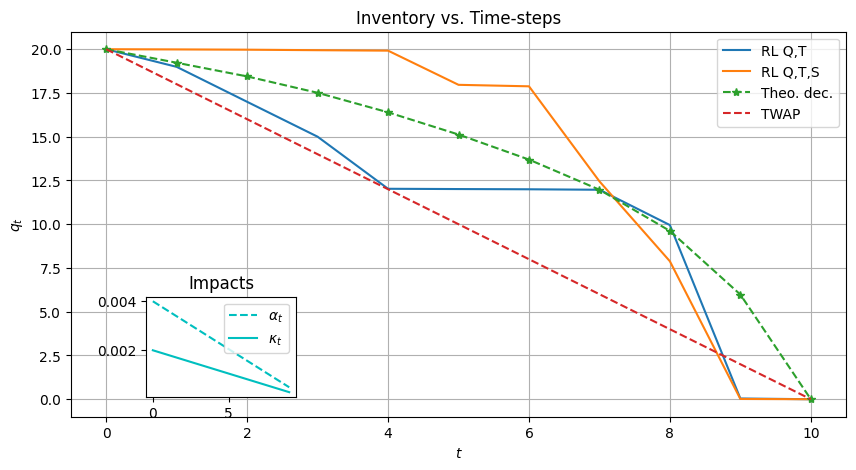}
    \caption{Average inventory holdings per time step with different features, compared to the optimal inventory holding if the strategy was known from the beginning, and to the inventory holding if the TWAP strategy was used}
    \label{fig:dec_inv}
\end{figure}

The impacts dynamics is $g(\frac{v_t}{\tau}) = \kappa_t\frac{v_t}{\tau} $, where $\kappa_t = \kappa_0 - \beta_{\kappa}\times t $ and $h(v_t) = \alpha_t v_t$,  where $\alpha_t = \alpha_0 - \beta_{\alpha} \times t$ and the parameters for the impacts are reported in Table~\ref{tab:pmts_tv_dn}.
 As before, the parameters are chosen such that the average value of the impact parameter during the trading period are equal to the fixed impacts in the experiments of the previous section. Fig. \ref{fig:dec_inv} shows the optimal solution for these parameters (green line, the red line is the TWAP reference). Since impacts are decreasing, it is optimal to trade less than in a TWAP at the beginning to avoid the initial high trading costs.
\par
\begin{figure}[ht]
    \centering
    
    \begin{subfigure}[ht]{0.5\textwidth}
        \centering
        \includegraphics[scale=0.45]{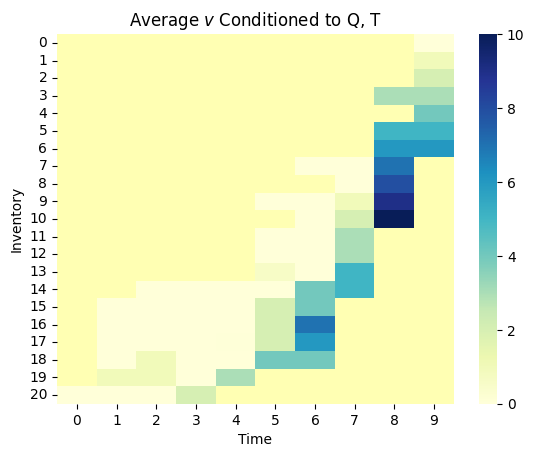}
    \end{subfigure}%
    \begin{subfigure}[ht]{0.5\textwidth}
        \centering
        \includegraphics[scale=0.6]{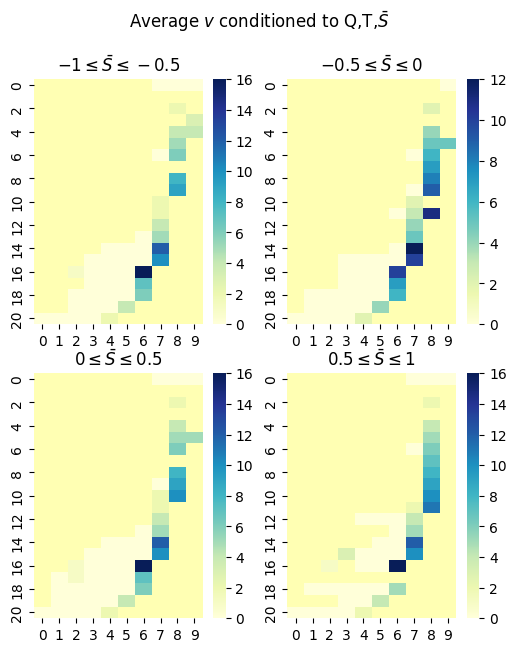}
    \end{subfigure}
    \caption{Decreasing impact. Left panel: Average number of shares sold per time-step $t$ and inventory level $q_t$. Right panel: Average number of shares sold per time-step $t$, inventory level $q_t$ and normalised price $\Bar{S}$}
    \label{fig:dec_both}
\end{figure}

\FloatBarrier
\begin{table}[htbp]
  \centering
  \caption{Parameters used in the decreasing time varying setting. The parameters are chosen such that the average value corresponds to the fixed impacts in the experiment above.}
  \label{tab:pmts_tv_dn}
    \begin{tabular}{||l|r||}
    \hline
    parameters      & value\\ 
    \hline  
    $\kappa_0$        & 0.002\\
    $\alpha_0$        & 0.004\\
    $\beta_{\kappa}$  & 0.0002\\
    $\beta_{\alpha}$  & 0.0004\\
    $\kappa_5=\kappa$ & 0.001\\
    $\alpha_5=\alpha$ & 0.002\\
    \hline
    \end{tabular}
\end{table}
\FloatBarrier
\paragraph{Features Q,T}
Similarly to the increasing impact case, when the features are inventory and time, the DDQL algorithm has a sub-optimal performance. In fact, as reported in Table~\ref{tab:tv_dec} 
its IS is higher than the theoretical optimum but the learnt strategy still outperforms the TWAP.
As in the previous case, the algorithm manages to understand the direction of the trend. As it can be seen in Fig.~\ref{fig:dec_inv} (blue line) and in the left panel of Fig.~\ref{fig:dec_both} the average actions per inventory level and time-step seem to be consistent with an increasing cost of execution, since the algorithm trades more toward the end of the period when costs are lower. It can be concluded that although correct in the estimation of the trend, the algorithm did not manage to calibrate its own actions to the magnitude of the impacts.
\paragraph{Features Q,T,S}
When including the mid-price among the features, the algorithm's performance gets fairly better. In fact, as it can be noticed in Table~\ref{tab:tv_dec}, when the algorithm features the mid-price does better than TWAP by lengths and performs slightly worse than the optimal trading strategy.
Again, adding the mid-price as a feature adds more power to the DDQL agent when it comes of estimating magnitude and direction of the impacts. This is confirmed by the average actions chosen per level of inventory, time step and normalised mid-price $\Bar{S}$, see Fig.~\ref{fig:dec_inv} (orange line) and the right panel of Fig.~\ref{fig:dec_both}. The strategy is again skewed to the right hand side of the heat-map implying that the algorithm has perfectly understood the direction and magnitude of the impacts.
\FloatBarrier
\begin{table}[htbp]
    \centering
    \begin{tabular}{ccccc|c}
        \hline
                        & \multicolumn{3}{c}{Decreasing}&  \\
                        & \textbf{Theo.} & \textbf{DDQL} & \Dpel vs. Theo.& \Dpel vs. Theo.  & \textbf{TWAP}\\
        \textbf{inputs} & $\Ex[IS]$  & $\Ex[IS]$ & \textbf{avg.}  & \textbf{std.dev.}& $\Ex[IS]$\\
        \hline
        \hline
        & & & & & \\
        Q,T & 0.2566    & 0.3080   & $  $-2.58 & $  $0.03 & 0.3588 \\
        & & & & & \\
        Q,T,S & 0.2566  & 0.2877  & $  $ -1.51 & $  $0.5 & 0.3588\\
        & & & & & \\
        \hline
    \end{tabular}
    \caption{Decreasing impact. Comparison between the costs of the DDQL agent, of the theoretical solutions, and of a TWAP strategy. $\Delta${P\&L} values are reported in basis points.}
    \label{tab:tv_dec}
\end{table}
\FloatBarrier

\FloatBarrier
\FloatBarrier
\FloatBarrier
\FloatBarrier
\subsubsection{Mixed train with increasing and decreasing impacts}

While the previous results were obtained by using the same dynamics in the train and test phase, here we consider a more realistic situation where the train phase contains different scenarios and in the test phase only one of them is actually realised. More specifically, the DDQL agent is now trained with $10,000$ trajectories using increasing impacts and $10,000$ trajectories using decreasing impacts. It is subsequently tested on $5,000$ trajectories
with either increasing or decreasing impacts trends. It is worth to study such experimental setup since it is functional for an eventual deploy in real market situations of an algorithm based on this RL paradigm. In fact, if the algorithm is used in a `production' setting, and especially without any prior knowledge of the data at hand, one would train such an algorithm with several scenarios for the stocks intended to be sold and then, using such strategy, one would like the algorithm to be able to detect the specific liquidity pattern and simultaneously adapt the trading strategy. We mimic this situation by training with both regimes, then we test with either of the two, in order to assess whether the DDQL agent is able to recognise the direction and the magnitude of the impacts at considered.

In this case the TWAP IS, used also above, becomes the natural benchmark. In fact, since the training set includes both increasing and decreasing impact while the test realises only one of them, the obvious choice for a agnostic trader would be to use a TWAP strategy. For this reason, in the tables we report also the \Dpel$_{TWAP}$ with respect to the TWAP strategy. The standard \Dpel is also reported but should be considered as a kind of unfeasible benchmark, since its calculation requires to know whether the impacts in the test set is increasing or decreasing.

\paragraph{Increasing impacts - Features Q,T}
Using as features only inventory and time, the agent is again not able to find the optimal strategy. As reported in Table~\ref{tab:tv_mix_inc} the performance in terms of \Dpel is still poor and the overall IS is similar to the one obtained with a TWAP strategy. As it can be seen in the left panel of Fig.~\ref{fig:mix_inc_qtp} the strategy achieved by the agent does not appear in line with the one that would be expected with such impact dynamics.
\paragraph{Increasing impacts - Features Q,T,S}
When the mid-price is included into the features, the performance of the DDQL agent is dramatically better. In fact, as reported in Table~\ref{tab:tv_mix_inc} the performance in terms of \Dpel is almost zero, meaning that the agent performs as good as the optimal solution in terms of costs. In turn, this implies that the agent has correctly estimated both the direction and the magnitude of the impacts used in the test phase. In fact, as it can be seen in the right panel of Fig.~\ref{fig:mix_inc_qtp}, the strategy found, conditioned to the price levels, is in line with the increasing magnitude in the impact dynamics.
\FloatBarrier
\begin{table}[htbp]
    \centering
    \begin{tabular}{ccccc|cc}
        \hline
                        & \multicolumn{3}{c}{Increasing Impact}& & \\
                        & \textbf{Theo.} & \textbf{DDQL} & \Dpel vs. Theo.& \Dpel vs. Theo.  & \textbf{TWAP}& \Dpel$_\textbf{TWAP}$\\
        \textbf{inputs} & $\Ex[IS]$  & $\Ex[IS]$ & \textbf{avg.}  & \textbf{std.dev.}& $\Ex[IS]$& \textbf{avg.} \\
        \hline
        \hline
        & & & & & &\\
        Q,T & 0.1449  & 0.2554   & $  $-5.34 & $  $ 1.1 & 0.2326 & {-0.92 }\\
        & & & & & &\\
        Q,T,S & 0.1449  & 0.1319  & $  $0.65  & $  $ 0.9& 0.2326& {5.2 }\\
        & & & & & &\\
        \hline
    \end{tabular}
    \caption{Increasing impact. Comparison between DDQL agent, theoretical solutions and TWAP strategy. $\Delta${P\&L} values are reported in basis points.}
    \label{tab:tv_mix_inc}
\end{table}
\FloatBarrier
\paragraph{Decreasing impacts - Features Q,T}
If just the inventory and the time-step form the state seen by the agent, the estimation of the parameters is noisy as in the other cases. Looking at Table~\ref{tab:mix_tv_dec} it can be seen that the resulting strategy performs slightly worse than the TWAP and it under-performs the optimal selling schedule suggested via optimisation of the cost functional. This result is in line with the strategy outlined in the left panel of Fig.~\ref{fig:mix_dec_qtp}.
\paragraph{Decreasing impacts - Features Q,T,S}
If the mid-price features the states seen by the agent the performance is fairly good if compared to the previous case. In fact, adding the price to the features of the algorithm increases the estimation power of DDQL and delivers a performance in line with the optimal one. This can be seen in Table~\ref{tab:mix_tv_dec} where the \Dpel is very low, suggesting that the impacts dynamics are estimated quite efficiently leading to a strategy consistent with the decreasing nature of the impacts' dynamics in the test set. This result is confirmed by the average action chosen per level of normalised mid-price, inventory and time-step shown in the right panel of Fig.~\ref{fig:mix_dec_qtp}.
\FloatBarrier
\begin{table}[htbp]
    \centering
    \begin{tabular}{ccccc|cc}
        \hline
                        & \multicolumn{3}{c}{Decreasing Impact}&  \\
                        & \textbf{Theo.} & \textbf{DDQL} & \Dpel vs. Theo.& \Dpel vs. Theo.  & \textbf{TWAP}&\Dpel$_\textbf{TWAP}$\\
        \textbf{inputs} & $\Ex[IS]$  & $\Ex[IS]$ & \textbf{avg.}  & \textbf{std.dev.}& $\Ex[IS]$&\textbf{avg.}\\
        \hline
        \hline
        & & & & & \\
        Q,T             & 0.2566    & 0.3696 & $  $-5.62  & $  $0.32 & 0.3588 & {-0.51 }\\
        & & & & & &\\
        Q,T,S           & 0.2566  & 0.2456  & $  $0.86  & $  $0.03 & 0.3588 & {6.5} \\
        & & & & & &\\
        \hline
    \end{tabular}
    \caption{Decreasing impact. Comparison between DDQL agent, theoretical solutions and TWAP strategy. $\Delta$P\&L values are reported in basis points.}
    \label{tab:mix_tv_dec}
\end{table}
\FloatBarrier
\subsection{Stochastic impacts}
Lastly, we consider stochastic impact coefficients. This experiment allows us to consider dynamics that are much more complicated and gives the opportunity to thoroughly test the capabilities of the DDQL agent to identify optimal strategies. As an analytical point of comparison, we use the third model of Section~\ref{financial models} for which an approximated solution
can be obtained as a Taylor expansion on the long run means of the temporary and permanent impact dynamics as outlined in Barger and Lorig (2019) \cite{barger2019optimal}. 
In this regime, the resulting optimal execution of Eq.~\eqref{actStoch} is in fact a perturbation of the TWAP strategy that accounts for deviation of the impacts with respect to their long run mean values.\par
The aim of the experiment is to test whether the agent is able to replicate, or even beat, the performance that would be obtained if the parameters for Eq.~\eqref{actStoch} were known. In fact, the difficulty in considering stochastic impacts when implementing an optimal execution strategy lies not just in the correct equation to be used, but in the correct parameters estimation for the impact model considered. By using the DDQL agent we aim at tackling the problem in a model agnostic way, i.e. we do not estimate or make assumptions on either the model or the impacts' magnitudes. We let the agent learn them via exploration-exploitation.\par  
The parameters chosen for the simulations are reported in Table~\ref{tab:stoch_pmts}. We assume that both impacts hover over their long run averages and are correlated, with correlation coefficient \footnote{This high value reflects the fact that different liquidity measures are typically highly correlated. We expect the same happens for temporary and permanent impact} $\omega = 0.9$. Thus, the parameters are chosen such that the Feller condition holds and the values for the impact coefficients are almost surely positive\footnote{{ It is worth noticing that in the space of solutions of \cite{barger2019optimal} both buys and sells are allowed for a sell program, while our RL approach considers only sells. In order to provide a fair comparison between the two methods, we selected the parameters of the model in such a way that, using the \cite{barger2019optimal} solution, the fraction of buys in a sell program is negligible. This parameter choice has been achieved via extensive numerical simulations.}}.
We implement two types of experiments, we test for strong mean reversion ($\lambda^{\text{(high)}}_\alpha$ and $\lambda^{\text{(high)}}_\kappa$) and for weak mean reversion ($\lambda^{\text{(low)}}_\alpha$ and $\lambda^{\text{(low)}}_\kappa$) parameters, leaving long run levels and volatility of the impacts untouched. Thus testing for either strong or weak memory properties in the dynamics for the impacts.

\begin{figure}[ht]
    \centering
    \begin{subfigure}[ht]{0.5\textwidth}
        \centering
        \includegraphics[scale=0.45]{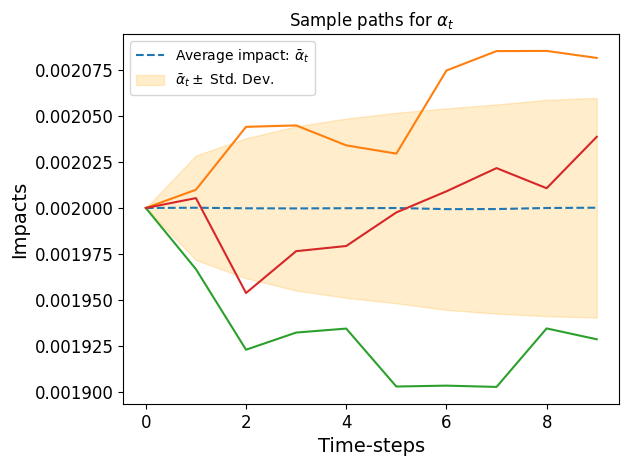}
    \end{subfigure}%
    \begin{subfigure}[ht]{0.5\textwidth}
        \centering
        \includegraphics[scale=0.45]{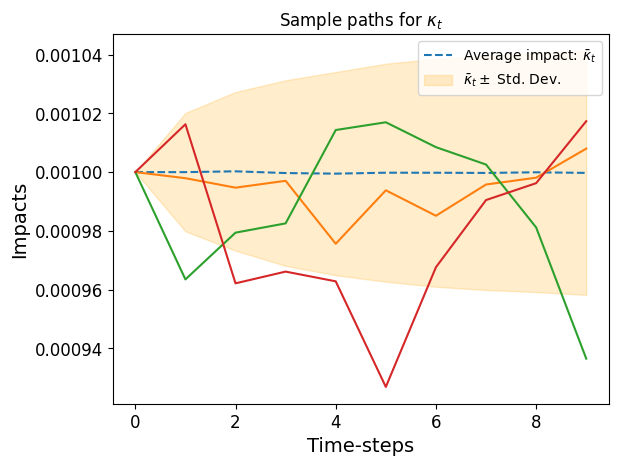}
    \end{subfigure}
    \caption{Stochastic impact. Left panel: three sample paths for $\alpha_t$ with $\lambda^{\text{(low)}}_\alpha$. Right panel: three sample paths for $\kappa_t$ with $\lambda^{\text{(low)}}_\kappa$.
    Solid lines: three different realisations for the temporary and permanent impacts. Yellow area is the standard deviation around the average value of the impact.}
    \begin{subfigure}[ht]{0.5\textwidth}
        \centering
        \includegraphics[scale=0.45]{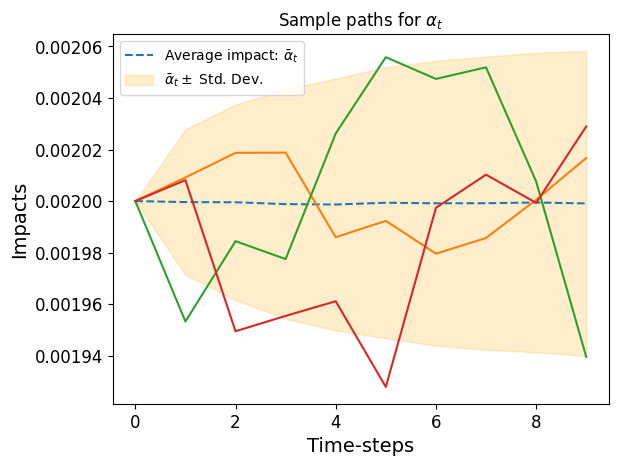}
    \end{subfigure}%
    \begin{subfigure}[ht]{0.5\textwidth}
        \centering
        \includegraphics[scale=0.45]{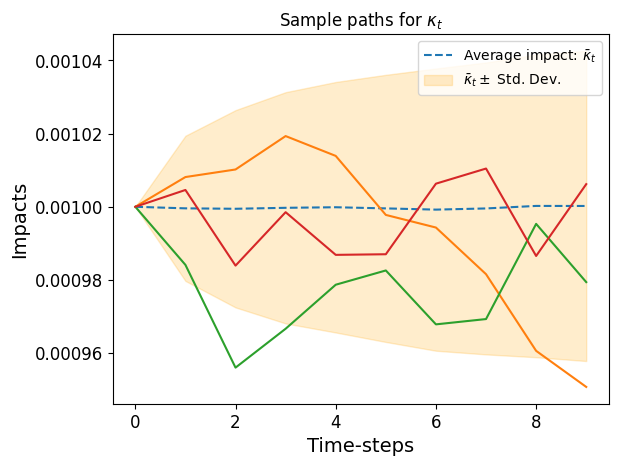}
    \end{subfigure}
    \caption{
    Left panel: sample paths for $\alpha_t$ with $\lambda^{\text{(high)}}_\alpha$. Right panel: sample paths for $\kappa_t$ $\lambda^{\text{(high)}}_\kappa$.
    Solid lines: three different realisations for the temporary and permanent impacts. Yellow area is the standard deviation around the average value of the impact.}
    \label{fig:a_tEk_t}
\end{figure}

In the experiments, for each train and test episode we simulate one path for $\alpha_t$ and one for $\kappa_t$. 
Sample paths for both $\alpha_t$ and $\kappa_t$ are reported in Fig~\ref{fig:a_tEk_t}.
\FloatBarrier
\begin{table}[htbp]
  \centering
  \caption{Parameters used for the simulation of stochastic impacts. The parameters are chosen such that the starting values correspond to the fixed impacts in the A\&C experiment above.}
  \label{tab:stoch_pmts}
    \begin{tabular}{||l|r||l|r||}
    \hline
    Parameter & Value & Parameter & Value \\ 
    \hline  
    $\lambda^{\text{(low)}}_\alpha $&$ 1$ & $\theta_\kappa$ & $0.001$ \\
    $\lambda^{\text{(low)}}_\kappa $&$ 1$ & $\theta_\alpha$ & $0.002$ \\
    $\lambda^{\text{(high)}}_\alpha $&$ 5$ & $\sigma_\kappa$ & $0.002$ \\
    $\lambda^{\text{(high)}}_\kappa $&$ 5$ & $\sigma_\alpha$ & $0.002$ \\
    $\omega$ & $0.9$ & & \\
    \hline
    \end{tabular}
\end{table}
\FloatBarrier
\FloatBarrier

\paragraph{Features - Q,T}
When the agent can only access the level of inventory and time-steps as features, when considering the $\lambda_{\alpha,\,\,\kappa}^{\text{(low)}}$ setup, the resulting IS is similar to the one obtained using Eq.~\eqref{actStoch} as shown in Table~\ref{tab:stoch_low}. The \Dpel is, in fact, slightly positive but less than 2 basis points. The performance increases in the case of $\lambda_{\alpha,\,\,\kappa}^{\text{(high)}}$ setup, as shown in Table~\ref{tab:stoch_high} it seems that since the impact dynamics mean revert faster to their long-run average the agent, trained in this way, takes advantage and trades accordingly, thus obtaining lower IS and better \Dpel.
  
\paragraph{Features - Q,T,S}
Adding the price to the set of features does bring significant benefits to the optimal selling schedule. Considering the  $\lambda_{\alpha,\,\,\kappa}^{\text{(low)}}$ setup, the \Dpel is now significantly positive, due to the increased amount of cash generated by training a DDQL agent with these features. Thus, adding the price to the set of features greatly improves the trading performance overall as seen in Table~\ref{tab:stoch_low}. When the $\lambda_{\alpha,\,\,\kappa}^{\text{(high)}}$ setup is considered, the performance of the agent is still extensively better than the approximated solution, but in line with those obtained with just inventory and time-step as features as shown in Table~\ref{tab:stoch_high}. We interpret this to be the result of the increased mean reversion rate, the agent learns that the dynamics tend to revert faster to the long run average and takes advantage with its trading.
Notably, the agent trained in this way has no notion of the stochastic dynamics of the impacts parameters used in this setting, the algorithm in fact, \textit{learns} the model and, implicitly, its parameters from the rewards obtained at each time-step and for each train episode. Thus, it is capable of performing better than an approximate closed-form solution, with the advantage that the parameters used to obtain the sale schedule are neither assumed nor estimated in any way, they are learned \textit{implicitly} and directly from the few features observed by the agent and fed to the neural nets.
\FloatBarrier
\begin{table}[htbp]
    \centering
    \begin{tabular}{ccccc|cc}
        \hline
                        & \multicolumn{4}{c}{Stochastic Impact: $\lambda_{\alpha,\,\,\kappa}^{\text{(low)}}$}&  \\
                        & \textbf{Theo.} & \textbf{Theo.}           & \textbf{DDQL} &  \textbf{DDQL} & \Dpel          & \Dpel \\
        \textbf{inputs} & $\Ex[IS]$      & $\textbf{std.dev.}[IS]$  & $\Ex[IS]$     &  $\textbf{std.dev.}[IS]$ & \textbf{avg.}  & \textbf{std.dev.}\\
        \hline
        \hline
        & & & & & &  \\
        Q,T             & 0.3129  & 0.63 & 0.2789 & 0.011 & 1.8 & 3.19  \\
        & & & & &  & \\
        Q,T,S &          0.3129   & 0.63 & 0.2572  & 0.007 &  2.5 & 3.27 \\
        & & & & &  &  \\
        \hline
    \end{tabular}
    \caption{Stochastic impact. Comparison between DDQL agent and theoretical solutions. $\Delta$P\&L values are reported in basis points.}
    \label{tab:stoch_low}
\end{table}
\FloatBarrier
\FloatBarrier
\begin{table}[htbp]
    \centering
    \begin{tabular}{ccccc|cc}
        \hline
                        & \multicolumn{4}{c}{Stochastic Impact: $\lambda_{\alpha,\,\,\kappa}^{\text{(high)}}$}&  \\
                        & \textbf{Theo.} & \textbf{Theo.}           & \textbf{DDQL} &  \textbf{DDQL} & \Dpel          & \Dpel \\
        \textbf{inputs} & $\Ex[IS]$      & $\textbf{std.dev.}[IS]$  & $\Ex[IS]$     &  $\textbf{std.dev.}[IS]$ & \textbf{avg.}  & \textbf{std.dev.}\\
        \hline
        \hline
        & & & & & &  \\
        Q,T             & 0.5017   & 1.83 & 0.3200 &  0.0084 & 9.2 &  73.3  \\
        & & & & & & \\
        Q,T,S &          0.5017    & 1.83 & 0.3158  &  0.0112 & 9.4 &   73.4 \\
        & & & & & &  \\
        \hline
    \end{tabular}
    \caption{Stochastic impact. Comparison between DDQL agent and theoretical solutions. $\Delta$P\&L values are reported in basis points.}
    \label{tab:stoch_high}
\end{table}
\FloatBarrier
\section{Conclusions}

In this paper we developed a DDQL algorithm able to learn the optimal liquidation strategy when the liquidity is time-varying. By considering an Almgren-Chriss linear market impact model, we investigated several deterministic and stochastic dynamics of the temporary and the permanent impact coefficient. These include constant, linearly varying, and square-root mean reverting stochastic processes for the parameters. We also investigate the case when coefficients are linearly varying but the slope can be either positive or negative with equal probability.

We find that, when considering settings where the optimal solution is known, the algorithm is able to learn strategies with an expected cost comparable with the optimal one. More interestingly, when we consider settings where the optimal solution is not available or is available in approximate form, we find that the DDQL algorithm finds better solutions. This indicates that RL could be successfully used in real settings where liquidity is naturally latent and time-varying. 

There are several possible extensions of our work. First of all, we considered a linear impact model, whereas it is known that temporary market impact is non-linear as shown in \cite{almgren2005direct}. Second, more sophisticated market impact models, such as the Transient Impact Model by \cite{bouchaud2003fluctuations} and its extension by \cite{bouchaud2009markets}, better describes the reaction of prices to trades. Finally, other dimensions of liquidity, e.g. the bid-ask spread, are important components of transaction cost and are known to be time varying with non-trivial dynamics. It is certainly interesting to explore if RL based approaches are capable to find optimal trading schedules also in these more complex and realistic environments. We leave these interesting questions for future research. 

\section*{Acknowledgements} F.L. acknowledges partial support by the European Program scheme ‘INFRAIA-01-2018-2019: Research and Innovation action’, grant agreement \#871042 ’SoBigData++: European Integrated Infrastructure for Social Mining and Big Data Analytics’. 

\printbibliography

@article{palmari2024,
title={Optimal execution in a time varying environment: well posedness and price manipulation},
  author={{Palmari, G.} and {Lillo, F.} and {Eisler, Z.}},
  journal={in preparation},
  year={2024}
}

@article{gatheral2012transient,
  title={Transient linear price impact and Fredholm integral equations},
  author={{Gatheral, J.} and {Schied, A.} and {Slynko, A.}},
  journal={Mathematical Finance: An International Journal of Mathematics, Statistics and Financial Economics},
  volume={22},
  number={3},
  pages={445--474},
  year={2012},
  publisher={Wiley Online Library}
}

@article{gueant2012optimal,
  title={Optimal portfolio liquidation with limit orders},
  author={{Gueant} and {Lehalle} and {Fernandez-Tapia}},
  journal={SIAM Journal on Financial Mathematics},
  volume={3},
  number={1},
  pages={740--764},
  year={2012},
  publisher={SIAM}
}

@article{almgren,
  title={Optimal execution of portfolio transactions},
  author={{Almgren, R.} and {Chriss, N.}} ,
  journal={Journal of Risk},
  volume={3},
  pages={5--39},
  year={2000}
}

@article{gueant2015general,
  title={General intensity shapes in optimal liquidation},
  author={Gu{\'e}ant, Olivier and Lehalle, Charles Albert},
  journal={Mathematical Finance},
  volume={25},
  number={3},
  pages={457--495},
  year={2015},
  publisher={Wiley Online Library}
}

@article{cartea2016incorporating,
  title={Incorporating order-flow into optimal execution},
  author={Cartea, Alvaro and Jaimungal, Sebastian},
  journal={Mathematics and Financial Economics},
  volume={10},
  pages={339--364},
  year={2016},
  publisher={Springer}
}

@article{casgrain2019trading,
  title={Trading algorithms with learning in latent alpha models},
  author={Casgrain, Philippe and Jaimungal, Sebastian},
  journal={Mathematical Finance},
  volume={29},
  number={3},
  pages={735--772},
  year={2019},
  publisher={Wiley Online Library}
}

@inproceedings{nevmyvaka2006reinforcement,
  title={Reinforcement learning for optimized trade execution},
  author={Nevmyvaka, Yuriy and Feng, Yi and Kearns, Michael},
  booktitle={Proceedings of the 23rd international conference on Machine learning},
  pages={673--680},
  year={2006}
}

@inproceedings{hendricks2014reinforcement,
  title={A reinforcement learning extension to the Almgren-Chriss framework for optimal trade execution},
  author={Hendricks, Dieter and Wilcox, Diane},
  booktitle={2014 IEEE Conference on Computational Intelligence for Financial Engineering \& Economics (CIFEr)},
  pages={457--464},
  year={2014},
  organization={IEEE}
}

@article{ning2021double,
  title={Double deep q-learning for optimal execution},
  author={Ning, Brian and Lin, Franco Ho Ting and Jaimungal, Sebastian},
  journal={Applied Mathematical Finance},
  volume={28},
  number={4},
  pages={361--380},
  year={2021},
  publisher={Taylor \& Francis}
}

@article{jeong2019improving,
  title={Improving financial trading decisions using deep Q-learning: Predicting the number of shares, action strategies, and transfer learning},
  author={Jeong, Gyeeun and Kim, Ha Young},
  journal={Expert Systems with Applications},
  volume={117},
  pages={125--138},
  year={2019},
  publisher={Elsevier}
}

@article{daberius2019deep,
  title={Deep execution-value and policy based reinforcement learning for trading and beating market benchmarks},
  author={Dab{\'e}rius, Kevin and Granat, Elvin and Karlsson, Patrik},
  journal={Available at SSRN 3374766},
  year={2019}
}

@article{sun2023reinforcement,
  title={Reinforcement learning for quantitative trading},
  author={Sun, Shuo and Wang, Rundong and An, Bo},
  journal={ACM Transactions on Intelligent Systems and Technology},
  volume={14},
  number={3},
  pages={1--29},
  year={2023},
  publisher={ACM New York, NY}
}

@article{hambly2021recent,
  title={Recent advances in reinforcement learning in finance},
  author={Hambly, Ben and Xu, Renyuan and Yang, Huining},
  journal={arXiv preprint arXiv:2112.04553},
  year={2021}
}

@article{abernethy2013adaptive,
  title={Adaptive market making via online learning},
  author={Abernethy, Jacob and Kale, Satyen},
  journal={Advances in Neural Information Processing Systems},
  volume={26},
  year={2013}
}

@inproceedings{kumar2020deep,
  title={Deep reinforcement learning for market making},
  author={Kumar, Pankaj},
  booktitle={Proceedings of the 19th International Conference on Autonomous Agents and MultiAgent Systems},
  pages={1892--1894},
  year={2020}
}

@article{yu2019model,
  title={Model-based deep reinforcement learning for dynamic portfolio optimization},
  author={Yu, Pengqian and Lee, Joon Sern and Kulyatin, Ilya and Shi, Zekun and Dasgupta, Sakyasingha},
  journal={arXiv preprint arXiv:1901.08740},
  year={2019}
}

@article{kearns2013machine,
  title={Machine learning for market microstructure and high frequency trading},
  author={Kearns, Michael and Nevmyvaka, Yuriy},
  journal={High Frequency Trading: New Realities for Traders, Markets, and Regulators},
  year={2013},
  publisher={Risk Books: London, UK}
}

@article{wei2019model,
  title={Model-based reinforcement learning for predictions and control for limit order books},
  author={Wei, Haoran and Wang, Yuanbo and Mangu, Lidia and Decker, Keith},
  journal={arXiv preprint arXiv:1910.03743},
  year={2019}
}

@article{briola2021deep,
  title={Deep reinforcement learning for active high frequency trading},
  author={Briola, Antonio and Turiel, Jeremy and Marcaccioli, Riccardo and Aste, Tomaso},
  journal={arXiv preprint arXiv:2101.07107},
  year={2021}
}

@inproceedings{karpe2020multi,
  title={Multi-agent reinforcement learning in a realistic limit order book market simulation},
  author={Karpe, Micha{\"e}l and Fang, Jin and Ma, Zhongyao and Wang, Chen},
  booktitle={Proceedings of the First ACM International Conference on AI in Finance},
  pages={1--7},
  year={2020}
}

@article{schnaubelt2022deep,
  title={Deep reinforcement learning for the optimal placement of cryptocurrency limit orders},
  author={Schnaubelt, Matthias},
  journal={European Journal of Operational Research},
  volume={296},
  number={3},
  pages={993--1006},
  year={2022},
  publisher={Elsevier}
}

@article{jaisson2022deep,
  title={Deep differentiable reinforcement learning and optimal trading},
  author={Jaisson, Thibault},
  journal={Quantitative Finance},
  volume={22},
  number={8},
  pages={1429--1443},
  year={2022},
  publisher={Taylor \& Francis}
}

@article{obizhaeva2013optimal,
  title={Optimal trading strategy and supply/demand dynamics},
  author={Obizhaeva, Anna A and Wang, Jiang},
  journal={Journal of Financial markets},
  volume={16},
  number={1},
  pages={1--32},
  year={2013},
  publisher={Elsevier}
}

@article{fouque2022optimal,
  title={Optimal trading with signals and stochastic price impact},
  author={Fouque, Jean-Pierre and Jaimungal, Sebastian and Saporito, Yuri F},
  journal={SIAM Journal on Financial Mathematics},
  volume={13},
  number={3},
  pages={944--968},
  year={2022},
  publisher={SIAM}
}

@article{mertens2022liquidity,
  title={Liquidity fluctuations and the latent dynamics of price impact},
  author={Mertens, Luca Philippe and Ciacci, Alberto and Lillo, Fabrizio and Livieri, Giulia},
  journal={Quantitative Finance},
  volume={22},
  number={1},
  pages={149--169},
  year={2022},
  publisher={Taylor \& Francis}
}

@article{bertsimas1998optimal,
  title={Optimal control of execution costs},
  author={Bertsimas, Dimitris and Lo, Andrew W},
  journal={Journal of financial markets},
  volume={1},
  number={1},
  pages={1--50},
  year={1998},
  publisher={Elsevier}
}

@article{almgren2005direct,
  title={Direct estimation of equity market impact},
  author={Almgren, Robert and Thum, Chee and Hauptmann, Emmanuel and Li, Hong},
  journal={Risk},
  volume={18},
  number={7},
  pages={58--62},
  year={2005}
}

@article{bouchaud2003fluctuations,
  title={Fluctuations and response in financial markets: the subtle nature ofrandom'price changes},
  author={Bouchaud, Jean-Philippe and Gefen, Yuval and Potters, Marc and Wyart, Matthieu},
  journal={Quantitative finance},
  volume={4},
  number={2},
  pages={176},
  year={2003},
  publisher={IOP Publishing}
}

@incollection{bouchaud2009markets,
  title={How markets slowly digest changes in supply and demand},
  author={Bouchaud, Jean-Philippe and Farmer, J Doyne and Lillo, Fabrizio},
  booktitle={Handbook of financial markets: dynamics and evolution},
  pages={57--160},
  year={2009},
  publisher={Elsevier}
}

@article{schied2010optimal,
  title={Optimal basket liquidation for CARA investors is deterministic},
  author={Schied, Alexander and Sch{\"o}neborn, Torsten and Tehranchi, Michael},
  journal={Applied Mathematical Finance},
  volume={17},
  number={6},
  pages={471--489},
  year={2010},
  publisher={Taylor \& Francis}
}

@book{cartea2015algorithmic,
  title={Algorithmic and high-frequency trading},
  author={Cartea, {\'A}lvaro and Jaimungal, Sebastian and Penalva, Jos{\'e}},
  year={2015},
  publisher={Cambridge University Press}
}

@article{sutton1998reinforcement,
  title={Reinforcement learning: an introduction MIT Press},
  author={Sutton, Richard S and Barto, Andrew G},
  journal={Cambridge, MA},
  volume={22447},
  year={1998}
}

@article{barger2019optimal,
  title={Optimal liquidation under stochastic price impact},
  author={Barger, Weston and Lorig, Matthew},
  journal={International Journal of Theoretical and Applied Finance},
  volume={22},
  number={02},
  pages={1850059},
  year={2019},
  publisher={World Scientific}
}

@article{campigli2022measuring,
  title={Measuring price impact and information content of trades in a time-varying setting},
  author={Campigli, F and Bormetti, G and Lillo, F},
  journal={arXiv e-prints},
  pages={arXiv:2212.12687},
  year={2022}
}
\newpage
\appendix
\section{Figures}
\FloatBarrier
\begin{figure}[ht]
    \centering
    \includegraphics[scale = 0.45]{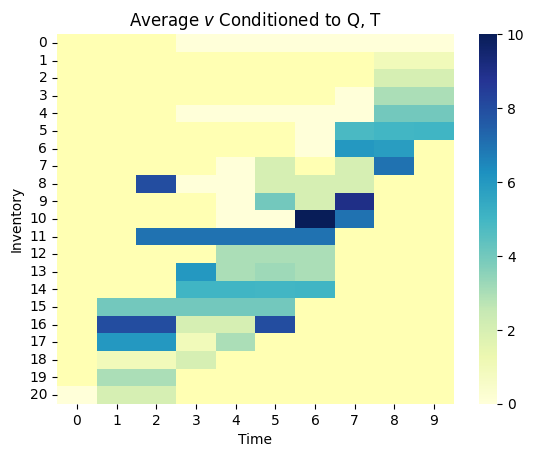}
    \label{fig:mix_dec_qt}
    \includegraphics[scale = 0.6]{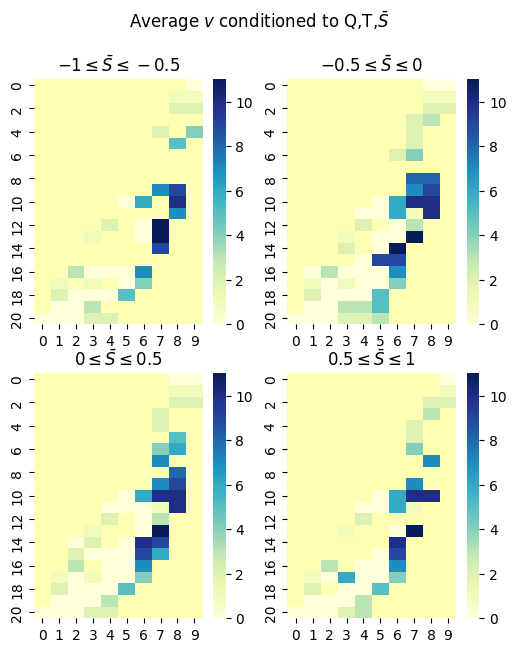}
    \caption{Mixed train, test with decreasing impact. Top panel: average number of stock sold per time-step $t$ and inventory level $q_t$. Bottom panel: average number of stock sold per time-step $t$, inventory level $q_t$ and normalised price $\Bar{S}$}
    \label{fig:mix_dec_qtp}
\end{figure}
\FloatBarrier
\begin{figure}
    \centering 
    \includegraphics[scale = 0.45]{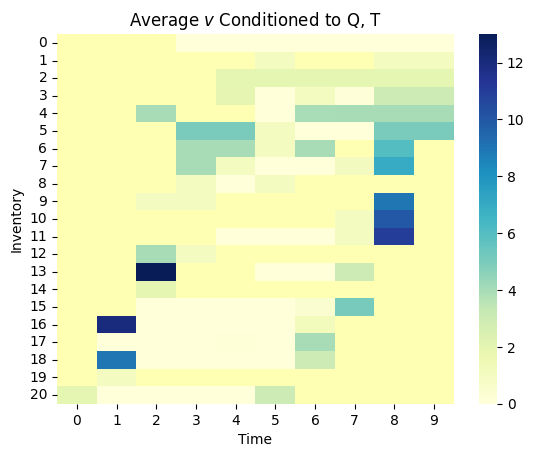}
    \label{fig:mix_inc_qt}
    \includegraphics[scale = 0.6]{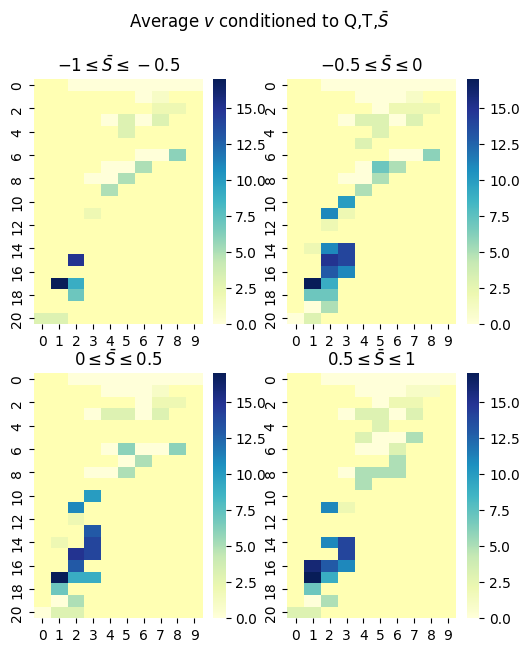}
    \caption{Mixed train, test with increasing impact. Top panel: average number of stock sold per time-step $t$ and inventory level $q_t$. Bottom panel: average number of stock sold per time-step $t$, inventory level $q_t$ and normalised price $\Bar{S}$}
    \label{fig:mix_inc_qtp}
\end{figure}
\FloatBarrier
\end{document}